\documentclass[aps,prl,reprint,showpacs,preprintnumbers,superscriptaddress, amsmath,amssymb,floatfix]{revtex4-2}

\usepackage{graphicx}% Include figure files
\usepackage{dcolumn}% Align table columns on decimal point
\usepackage{bm}% bold math
\usepackage{hyperref}% add hypertext capabilities
\usepackage[normalem]{ulem}

\usepackage{color}
\usepackage{subfigure}
\definecolor{niceblue}{rgb}{0.388235, 0.627451, 0.847059}
\definecolor{nicered}{rgb}{0.7,0.1,0.1}
\definecolor{nicegreen}{rgb}{0.1,0.5,0.1}

\begin{document}

\title{LC Circuits for the Direct Detection of 
Ultralight Dark Matter Candidates}

\author{Christopher~M.~Donohue}
\email{cmd298@cornell.edu}
\affiliation{Department of Physics, Cornell University, Ithaca, NY 14853-0001, USA} 
\affiliation{Department of Physics and Astronomy, University of Kentucky, Lexington, KY 40506-0055, USA}
\author{Susan~Gardner}
\email{gardner@pa.uky.edu}
\affiliation{Department of Physics and Astronomy, University of Kentucky, Lexington, KY 40506-0055, USA}
\author{Wolfgang~Korsch}
\email{Wolfgang.Korsch@uky.edu}
\affiliation{Department of Physics and Astronomy, University of Kentucky, Lexington, KY 40506-0055, USA}


\begin{abstract}
Cosmological mechanisms that yield ultralight dark matter are insensitive to the intrinsic parity of a bosonic dark matter candidate, but that same quantity plays
a crucial role in a direct detection experiment. 
The modification of electrodynamics in the presence of ultralight axion-like dark matter
is well-known and has been used	to realize sensitive probes of such sub-eV mass-scale
dark matter, and analogous studies exist for hidden-photon dark matter as well. 
Here we reframe	the modification of electrodynamics for ultralight dark matter
of positive intrinsic parity, with a 
focus on the scalar case. 
In particular, we show that 
resonant LC circuit searches for axions can be modified
to detect scalar dark matter particles by exploiting the large electric fields
developed for use in neutron EDM experiments. Our proposed experimental set-up can improve upon
previous sensitive searches 
for scalar particles from ``light shining
through a wall'' experiments to probe scalar-photon couplings some three orders of
magnitude smaller in the $1\times 10^{-11} - \,4\times 10^{-8}$ eV mass 
($2\, {\rm kHz} - 10\,{ \rm MHz}$ frequency) range. 
\end{abstract}

\maketitle

{\it Introduction.} 
Despite the preponderance of astrophysical evidence in support of 
the existence of dark matter, its 
essential nature has remained elusive. Although the dark matter
candidates motivated by weak-scale supersymmetry, the WIMP~\cite{Jungman:1995susyDM}, and by 
the explanation of why the strong interaction does not break P and CP 
symmetries, the axion~\cite{PecceiQuinn1977PhRvL..38.1440P,PecceiQuinn1977PhRvD..16.1791P,Weinberg1978newlightboson,Wilczek1978PhRvL..40..279W}, remain well-motivated, it has become 
apparent that yet broader possibilities exist and can act as alternative solutions
to the dark matter problem~\cite{Feng:2010DMreview}. In particular, a particle dark matter candidate 
could be much heavier or much lighter in mass than the ${\cal O}(100\,\rm keV)$ (``visible'') axion
and the ${\rm MeV}$-$100\, \rm TeV$ WIMP mass 
range suggested by theory, and in which many searches have been made.
The 
``visible'' axion is regarded as excluded~\cite{Dolan2017revisedaxionlimits}, although 
the existing experimental constraints can be evaded~\cite{Alves2017viableQCDaxion}, and both heavier and lighter axions are also possible and are the targets of experimental searches~\cite{Dolan2017revisedaxionlimits}. 
It is the purpose of this article to explore the possibility of sub-eV dark 
matter more 
broadly and to show that this larger set of 
possibilities can be probed with existing technology. 

The possibility of sub-keV mass 
particle dark matter candidates was first
explored in the context of axion  cosmology~\cite{Preskill1983cosmologyaxion,Abbott1983cosmologicalaxion,Dine1983notsoharmlessaxion}, in which it was realized, although thermally cold axions could be 
readily produced in the early Universe 
through the QCD phase transition, that if they were too light in mass, say 
less than some 10 $\mu$eV, too much dark matter would be produced, 
making the scenario incompatible with the Universe as it is observed. 
Although the ${\cal O}(10\,\mu \rm eV)$ mass range is a highly motivated 
search window~\cite{ADMX:2020run1Bresults}, 
still broader mass ranges become possible in particular
cosmological histories with inflation~\cite{Linde1988inflationaxion}. 
For example, the axion is associated with the spontaneous breaking of 
Peccei-Quinn symmetry~\cite{PecceiQuinn1977PhRvL..38.1440P,PecceiQuinn1977PhRvD..16.1791P}, and the relationship between 
its mass and its coupling to light is 
essentially 
fixed once the energy scale $f_a$ at which the symmetry is
broken is known, where we note the KSVZ~\cite{Kim1979KSVZ,Shifman1979KSVZ} and DFSZ~\cite{Dine1981DFSZ,Zhitnitsky1980DFSZ} ``invisible'' axion models. 
If $f_a$ exceeds the energy scale of inflation, then
the ``extra'' matter can be inflated away and no longer contribute to the
mass of the Universe as we observe it. Thus the viability of sub-10 $\mu$eV 
dark matter is tied to the cosmological history of our Universe~\cite{Linde1988inflationaxion}, opening many more 
possibilities. 
Indeed, if the axion's mass and coupling to light are regarded
as independent parameters, making the particle, rather, axion-like, 
then the possible parameter space
becomes enormously 
broader~\cite{Graham2011axionDMcoldmolecule,Graham2013newdirectdetectaxion}.
In what follows we refer to particle dark matter with sub-eV mass as 
ultralight. 

The properties of ultralight dark matter 
are qualitatively
different from those of WIMP-like candidates, because their associated number
densities are grossly different. 
We recall that cosmological simulations of the evolution of Milky-Way-like galaxies, 
as well as astrometric observations of the Milky Way itself, 
reveal the local dark matter density to be roughly $0.3\,\rm GeV/cm^3$~\cite{Gardner2021ppnpreview}, 
so that the number density of sub-eV dark matter candidates is enormous, making their
behavior wave-like and essentially quantum mechanical in nature. 
The ubiquity of the Pauli Principle implies that such dark matter must
be bosonic in nature~\cite{Baldeschi1983fermionvsbosonDMhalo}, 
because the Pauli repulsion between such 
ultralight fermions would make them no longer bound to the galaxy in 
which they reside, in conflict with the observed galactic rotation curves. 
Moreover, its minimal possible particle mass is about
$10^{-22}\,\rm eV$~\cite{Gardner2021ppnpreview}, as yet lighter mass
dark matter would not ``fit'' into dwarf 
galaxies~\cite{Hui2017ultralightscalarDM}.
Thus far 
axion-like and hidden-photon 
dark matter candidates have
been considered, 
in favor of yet higher spin candidates, 
and the production mechanism need not be
tied to the QCD phase transition. Rather, a misalignment 
mechanism~\cite{Graham2011axionDMcoldmolecule,Graham2013newdirectdetectaxion,Nelson2011vectorDMinflation,Arias2012wispycdm} is possible.
After inflation, the light field can take a random nonzero value in a casually connected
region of the Universe and oscillations in that field can be interpreted
as particles~\cite{Nelson2011vectorDMinflation}.
We emphasize that these cosmological scenarios in no way select
the parity of the dark matter particle, so that in addition to 
axion and hidden photon dark matter candidates, which each have negative
intrinsic parity, dark matter candidates with positive intrinsic
parity are possible as well. This is important because if parity
symmetry is not broken, then the coupling of opposite intrinsic parity
particles to light is quite different. Here we explore the consequences of
these differences to realize new ways of detecting ultralight dark matter. 

{\it Theoretical Framework.}
Axion electrodynamics~\cite{Sikivie1983invisibleaxiontests,Wilczek1987axioelectrodynamics} is the 
extension of electrodynamics to include interactions with the axion. 
To realize ultralight hidden-photon dark matter, 
the hidden photon is given a Stueckelberg mass~\cite{Nelson2011vectorDMinflation}; and 
electrodynamics is modified through the kinetic mixing of $F^{\mu\nu}$ with a hidden photon 
tensor $F'^{\mu\nu}$, with detectable consequences in the sub-$\mu$eV 
regime~\cite{Chaudhuri2015hiddenphoton_dmradio}. 
Here we consider just one type 
of dark matter candidate at a time and assume that parity is conserved. To segue from negative to 
positive intrinsic parity dark matter candidates and working in MKS units, we replace 
\begin{equation}
{\cal L}_{0^-} \supset -\frac{1}{4\mu_0} g_a a F_{\mu\nu} \widetilde{F}^{\mu\nu} \rightarrow
{\cal L}_{0^+} \supset -\frac{1}{4\mu_0} g_\phi \phi F_{\mu\nu} {F}^{\mu\nu} 
\end{equation}
and 
\begin{equation}
{\cal L}_{1^-} \supset -\frac{\varepsilon_v}{2\mu_0} F_{\mu\nu} {F'}^{\mu\nu} \rightarrow
{\cal L}_{1^+} \supset -\frac{
\varepsilon_a}{2\mu_0} F_{\mu\nu} \widetilde{F'}^{\mu\nu} \,.
\end{equation}
Thus in the spin-zero sector, 
to account for the different intrinsic parity, a electromagnetic field tensor, $F^{\mu\nu}$, needs to replace the dual tensor, $\widetilde{F}^{\mu\nu}=
\epsilon^{\mu\nu\alpha\beta}F_{\alpha\beta}/2$, in the axion-electromagnetic interaction, 
with an analogous replacement in the spin-one case. 
Finally the interaction term between the scalar and electromagnetic field becomes 
\begin{equation}
    \mathcal{L}_{int}=-\frac{1}{4\mu_0}g_\phi \phi(x) F_{\mu\nu}F^{\mu\nu}=\frac{g_\phi}{2\mu_0}\phi(x)(\vec{E}^2-\vec{B}^2) \,.
\end{equation}
Here, $\phi(x)$ is the scalar field, 
$g_\phi$ is the scalar-photon coupling constant, and $\vec{E}$ and $\vec{B}$ are the electric and magnetic fields, respectively.

The coupling of the scalar field to electromagnetism implies that the inhomogeneous Maxwell equations are modified through ${\cal O}(g)$
as follows:
\begin{equation}
    \Vec{\nabla}\cdot \Vec{E}=g_\phi\Vec{\nabla}\phi\cdot\Vec{E}+\frac{\rho_{e}}{\varepsilon_o} \,;
\end{equation}
\begin{equation}\label{eq:5}
    \Vec{\nabla}\times\Vec{B}-\frac{1}{c^2}\frac{\partial \Vec{E}}{\partial t}=g_\phi\left(\Vec{\nabla}\phi\times\Vec{B}-\frac{1}{c^2}\Vec{E}\frac{\partial \phi}{\partial t}\right)+\mu_0\Vec{j}_{e}\,,
\end{equation}
where $\rho_{e}$ and $\Vec{j}_{e}$ are the electric charge and current densities associated with ordinary electromagnetism.

Just as for the axion field~\cite{Sikivie1983invisibleaxiontests}, we assume 
the potential of the scalar field to be 
of simple harmonic form, so that 
$U_\phi=m_{\phi}^2\phi^2 / 2$. 
This potential and the interaction term imply the wave equation:
\begin{equation}\label{eq:6}
    \Box\phi=\frac{2g_\phi}{\mu_0}(E^2-B^2)-m_{\phi}^2\phi \,.
\end{equation}
If $\vec{E}$ and $\vec{B}$ are static, 
then the scalar field oscillates with angular frequency:
\begin{equation}
    \omega= \frac{m_\phi c^2}{\hbar}(1+\frac{1}{2}\vec{v}^2) \,,
\end{equation}
where $\vec{v}$ is the scalar field velocity, in the laboratory rest frame. 

{\it Existing Constraints.} The possibility of ultralight scalar dark matter 
can be probed in a similar manner to that of 
a weakly coupled
axion~\cite{Sikivie1983invisibleaxiontests}, as through tests of
electrodynamics~\cite{Okun1982limitselectrodynamics}. 
Particular constraints derive from searches for 
(i) the 
possibility of ``light shining through a wall'' (LSW)~\cite{Okun1982limitselectrodynamics,Sikivie1983invisibleaxiontests,vanBibber1987psthroughwalls}, 
(ii) the appearance of new, macroscopic
interactions~\cite{MoodyWilczek1984newforces,Adelberger2009torsionbalancerev}, or (iii) the 
lack of anomalous energy loss from stars, stellar remnants, or 
supernovae~\cite{Raffelt2008astroaxionbounds}. 
We note that the limits on new, macroscopic forces~\cite{Leefer2016fifthforcescalars} 
directly limit scalar-fermion couplings only~\cite{MoodyWilczek1984newforces}.
LSW experiments can be conducted in either  
broadband~\cite{Ehret2010LSW_ALPS,Ballou2015_OSQAR} 
or resonant versions~\cite{Betz2013LSW_CROWS}.
For sub-eV scalar dark matter, the most severe limits on scalar-photon
couplings come from the LSW experiment of Ballou et al.~\cite{Ballou2015_OSQAR} and, for $m_\phi \lesssim 10^{-15}\, \rm eV$, 
from searches for the direct impact of such relic dark matter on 
atomic spectroscopy~\cite{VanTilburg2015scalardmatomicspect,Hees2016massivescalarHF}
and 
atomic clocks~\cite{Stadnik2016PhRvAimprovedlimitsscalaratomicclock}. 

{\it Direct Detection via an LC circuit.}
As Eq.~\eqref{eq:5} shows, a substantial difference between the dynamics of the scalar and pseudoscalar cases is that the time-dependent portion of the dark matter-generated current 
depends on an electric field instead of a magnetic field. Thus detection methods developed
for axion-like dark matter can be adapted to the scalar case, 
where we note the LSW experiment of Ref.~\cite{Ballou2015_OSQAR} as 
an existing example. 
The possibility of powerful empirical constraints on the scalar-photon
coupling $g_\phi$ arise if the 
scalar-generated magnetic field is amplified 
using a resonant LC circuit and detected with a 
SQUID magnetometer.
Consequently, this leads to a natural adaptation of the proposal of  Ref.~\cite{Sikivie2014LCcircuit}; namely, the use of a large static electric field,
rather than a large magnetic field, 
to produce a scalar-generated magnetic field, with a similar expected
enhancement in a 
superconducting LC circuit. Yet further 
experiments for ultralight axion or hidden photon dark 
matter are under development~\cite{Ouellet2018abracadabra,Chaudhuri2015hiddenphoton_dmradio,Berlin2020heterodyneaxion} and could potentially 
be reframed to consider
the even intrinsic parity dark matter candidates of interest to us here. 
 
\begin{figure}
    \centering
    \includegraphics[scale=.45]{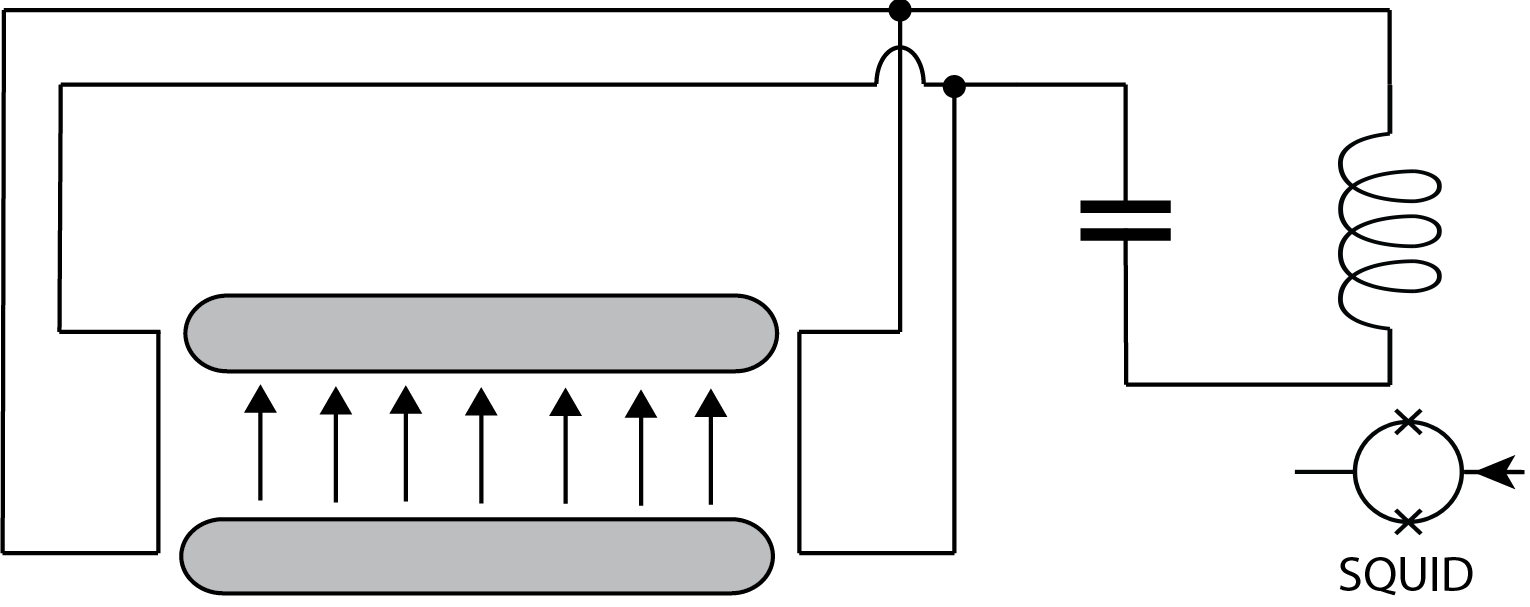}
    \caption{Schematic drawing of the proposed scalar dark matter detector. The grey rectangles are electrodes. The pick-up loops are connected to an LC circuit coupled with a SQUID magnetometer. The field $\vec{E}_o$ is represented by the arrows between the electrodes.
    }
    \label{fig:LC Circuit}
\end{figure}

If the spatial extent of a static electric field is less than $\sim m_\phi^{-1}$ 
and the spatial gradient in $\phi$ is assumed negligible,
then $\partial\vec{E}/\partial t=0$ and $\Vec{\nabla}\phi\times\Vec{B}\approx 0$ in Eq.~\eqref{eq:5}.   Therefore, in the presence of a static, uniform electric field, $\vec{E}_o$, Eq.~\eqref{eq:5}, becomes:
\begin{equation}\label{eq:8}
    \vec{\nabla}\times \vec{B}_{\phi}=-\frac{g_\phi}{c^2}\vec{E}_o\frac{\partial\phi}{\partial t}\equiv\vec{j}_\phi \,,
\end{equation}
where $\vec{j}_\phi$ is the dark matter produced electric current. As in Ref.~\cite{Sikivie2014LCcircuit}, pick-up loops connected to an LC circuit can amplify this magnetic field when the resonant frequency of the circuit, 
$\omega/2\pi=1/\sqrt{LC}$, is near $m_\phi c^2/h$. 
Thereby, both the coupling constant, $g_\phi$, and the particle mass, $m_\phi$, can be detected. However, as the schematic drawing of the proposed experimental set-up in Fig.~\ref{fig:LC Circuit} shows, the superconducting pick-up loops are placed outside the electric field. This prevents voltage breakdown due to the pick-up loops. 
Additionally, multiple loops 
may be placed around a circular electric field in order to increase the total magnetic flux through the LC circuit. When the resonant frequency of the LC circuit equals that of the scalar mass, since the wire is superconducting, the current caused by $\vec{B}_\phi$ equals:
\begin{equation}
    I=-Q\Phi_\phi/L \,,
\end{equation}
where $Q$ is the quality factor of the circuit, $\Phi_\phi$ is the magnetic flux of $\vec{B}_\phi$ through the pickup loops, and $L$ is the inductance of the circuit in its environment. This leads to a magnetic field detected by the SQUID magnetometer:
\begin{equation}\label{eq:9}
    B_d\approx \mu_0\frac{N_d}{2r_d}I=-\mu_0\frac{QN_d}{2Lr_d}\Phi_\phi \,,
\end{equation}
where $N_d$ is the number of turns and $r_d$ is the radius of the inductor facing the magnetometer.
If the radius of the electrodes is $r_e$ with a separation 
distance $d$, then using Eq.~\eqref{eq:8}, and Stokes' Theorem, the scalar-generated magnetic field outside the electric field is:

\begin{equation}
    \vec{B}_\phi=\frac{g_\phi}{2 c^2}E_o\frac{\partial \phi}{\partial t} r_e^2 \frac{1}{r}\hat{\varphi} \,,
\end{equation}
where $(z,r,\varphi)$ are cylindrical coordinates centered at the middle of the electrodes, with
$\varphi>0$ in the counterclockwise direction.
If the pick-up loops have length $l$ and are of the same height as the electrode separation distance, then the magnetic flux through a single loop is:
\begin{equation}\label{eq:10}
    \Phi_\phi=-\frac{1}{2c^2}g_\phi E_o\frac{\partial \phi}{\partial t} r_e^2d\ln{(1+l/r_e)} \,.
\end{equation}
For multiple loops, this quantity would be multiplied by the number of loops, $N_\ell$. 

If the scalar field is real, as is considered for the axion field in Ref.~\cite{Foster2017axiondirectdetection}, then the dark matter energy density is related to the time derivative of the scalar through:
\begin{equation}\label{eq:11}
    \rho_{DM}=\frac{1}{\hbar c^3}\left(\frac{\partial \phi}{\partial t}\right)^2 \,.
\end{equation}
Combining Eqs.~\eqref{eq:9}, \eqref{eq:10}, and \eqref{eq:11}, the detected magnetic field would be:
\begin{align}
  &B_d \approx
  \mu_0\sqrt{\frac{\hbar}{c}}\frac{QN_d N_\ell}{4Lr_d}g_\phi E_o\sqrt{\rho_{DM}} r_e^2d\ln{(1+l/r_e)}
  \nonumber\\
 &=
    1.47\times 10^{-20}\rm{T} \left(\frac{Q}{10^4}\right) N_d N_\ell \left(\frac{\rm{\mu H}}{L}\right)\left(\frac{cm}{r_d}\right)\left(\frac{r_e^2d}{\rm{cm}^3}\right)
    \nonumber\\
    &\times\!\left(\frac{g_\phi}{10^{-12}\,{\rm GeV}^{-1}}\right)\!
    \left(\frac{E_o}{\rm{10\,kV/cm}}\right)\!
    \sqrt{\frac{\rho_{DM}}{\rm{GeV/cm^3}}}\! \ln{(1+l/r_e)}\,.
\end{align}

\begin{figure}
    \centering
    \includegraphics[scale=.60]{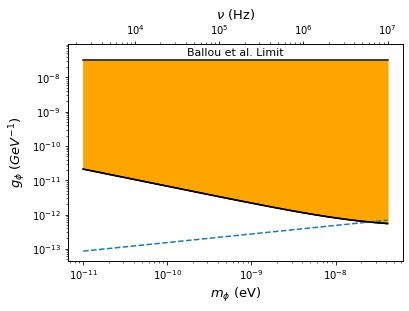}
    \caption{Expected sensitivity for our proposed experiment. 
    The upper bound of the shaded area is set by Ballou et al.~\cite{Ballou2015_OSQAR}. The lower bound is set by $S/N=5$. The dashed blue line shows the possible values of $g_\phi$ which can be detected due to the expected sensitivity of the magnetometer; we refer to the text for additional discussion.}
    \label{fig:sensitivty}
\end{figure}

{\it Sensitivity Estimates.} The sensitivity for the experiment is determined by the signal 
to noise ratio $S/N$ for the 
current, where we note Ref.~\cite{Foster2017axiondirectdetection} for a comparison of 
this criterion with a likelihood analysis, as well as by the sensitivity of the SQUID magnetometer. Following 
Ref.~\cite{Sikivie2014LCcircuit}, 
we assume the loci of parameters $g_\phi$ and $m_\phi$ that would generate a 
$S/N>5$ would be detectable and thus could be excluded. 
The signal due to the induced current is:
\begin{align}
\label{signal}
    \!\!\!\!\!\!&I=\sqrt{\frac{\hbar}{c}}\frac{Q}{2L}N_\ell g_\phi E_o\sqrt{\rho_{DM}}r_e^2 d\ln{(1+l/r_e)}
    \nonumber\\
    &=2.34\times 10^{-16}\rm{A} \left(\frac{Q}{10^4}\right)\left(\frac{\rm{\mu H}}{L}\right)N_\ell\left(\frac{g_\phi}{\rm{10^{-12}\,GeV^{-1}}}\right)\nonumber\\
    &\times \left(\frac{E_o}{10\, \rm{kV/cm}}\right)\sqrt{\frac{\rho_{DM}}{\rm{GeV/cm^3}}}\left(\frac{r_e^2 d}{{\rm cm}^3}\right)\ln{(1+l/r_e}) \,.
\end{align}
The expected main sources of noise in its detection
are the Johnson-Nyquist thermal noise 
$\delta I_T$ at temperature $T$ with circuit noise $\Delta \nu$ and the 
noise $\delta I_B$ associated with that in the magnetometer $\delta B$, 
namely~\cite{Sikivie2014LCcircuit}, 
\begin{align}
\label{Tnoise}
    \!\!\!\!\!\!&\delta I_T=\sqrt{\frac{4 k_B T Q \Delta \nu}{L\omega}}
            =2.96\times 10^{-13} \rm{A} \nonumber\\
    &\times \sqrt{
    \left(\frac{{\rm MHz}}{\nu}\right)
    \left(\frac{Q}{10^4}\right) 
    \left(\frac{\mu{\rm H}}{L}\right) 
    \left(\frac{T}{{\rm mK}}\right) 
    \left(\frac{\Delta \nu}{\rm{mHz}}\right)
    }
    \end{align}
and
\begin{align}
\label{Bnoise}
    \!\!\!\!\!\!&\delta I_B = \frac{ 2r_d}{N_d} \delta B = 5.03 \times 10^{-14} \rm{A} 
    \nonumber\\
    &\times \frac{1}{N_d} 
    \left( \frac{r_d}{{\rm cm}}\right)
    \left(\frac{B_n}{10^{-16} {\rm T}}\right) 
    \sqrt{\left(\frac{\Delta \nu}{\rm{mHz}}\right)}
     \,,
\end{align}
where $\delta B = B_n \sqrt{\Delta \nu/{\rm Hz}}$ and $B_n \approx 10^{-16}\,{\rm T}$, 
with the former noise source being numerically larger. 
We note that 
additional RF noise was noticed in its experimental realization~\cite{ADMX-SLIC2020}.
In the current case, additional, possible 
sources of noise or false signals come from 
non-uniformity or drift in the electric field. If the field is not uniform, the signal could be weakened since the scalar-generated magnetic field might not be perfectly symmetric around the electrodes under these circumstances. Also, if 
the electric field drifts, due to the $\partial\vec{E}/\partial t$ in Eq.~\eqref{eq:5}, it could 
potentially result in a magnetic field comparable to that of the dark matter signal. 
Finally, if $\partial\vec{E}/\partial t$ is greater than $g_\phi E_o\partial \phi/\partial t$, then the signal from the dark matter can be drowned out. 
Spatial and/or temporal variations
in the ambient magnetic field could also prove to be an important source of systematic error, though the experiment should probably be conducted in a magnetically shielded environment.
Nevertheless, for concrete comparison with the sensitivity 
of the axion dark-matter experiment proposed in 
Ref.~\cite{Sikivie2014LCcircuit}, we compute
\begin{equation}
    S/N=\frac{I}{\sqrt{(\delta I_T)^2+(\delta I_B)^2}} \,,
\end{equation}
noting, too,
that for a superconducting circuit $L=N_\ell L_m+L_c + L_d$, with the inductance coming for 
a pickup coil, the coaxial cable, and the inductor facing the SQUID magnetometer, respectively.
Here $L_d = N_d^2 \xi$ with $N_d$ coils and $\xi=\mu_0 r_d({\rm ln}(8r_d/a_d)-2)$\footnote{The 
self-inductance $\xi$ is the 
high-frequency limit of a circular ring of radius $r_d$ of a wire of circular section
with radius $a_d$, neglecting terms of ${\cal O}(a_d^2/r_d^2)$, as per Eq.~(62) of 
Ref.~\cite{rosa1912formulas}.}.
We eschew the
optimization procedure of Ref.~\cite{Sikivie2014LCcircuit} for $N_d$ because $S/N$ depends on
a plurality of inputs and $S/N$ varies slowly with $N_d$. 
We emphasize that 
the sensitivity of the experiment can be limited by 
the sensitivity of the magnetometer. If the magnetic field generated by $\phi$ is so small 
that the magnetometer cannot detect it, then even if the signal to noise ratio is still large enough the signal will not  be detected. In the few hundred kHz frequency
regime we study here, magnetometer sensitivities of $10^{-16}\,\rm T$ have been 
established~\cite{Koerber2016SQUID_SuScT..29k3001K}, with increases of sensitivities
possible from the coherence of the dark matter field and its integration time. 
Under the assumption of the standard dark matter halo model~\cite{Gardner2021ppnpreview},
the energy density is $\rho_{DM}\approx 0.3\, \rm{GeV/cm^3}$, and we suppose 
the coherence time 
is $t_c\approx 0.16\, {\rm s}\, (\rm{MHz}/\nu)$~\cite{Sikivie2014LCcircuit}. 
The magnetometer can detect a field: $B_d=10^{-16}{\rm T} ({\rm Hz})^{-1/2}(tt_c)^{-1/4}$ with $t$ being the integration time. If the integration time is $t=10^3\rm{s}$, then $B_d\approx 2.8\times 10^{-17}\rm{T} (\nu/\rm{MHz})^{1/4}$~\cite{Sikivie2014LCcircuit}. We show the coupling associated with
the smallest detectable magnetic field in 
Fig.~\ref{fig:sensitivty}, illustrating, for larger values of $m_\phi$, that 
this factor can limit the sensitivity of the proposed experiment. 
If we were to employ the caustic ring halo model~\cite{Sikivie2008CausticRing} instead, the coherence time would be longer, making the detection of still smaller $B_d$ possible. 
For this article, though, we focus on the standard halo model. 

LSW experiments from the OSQAR collaboration have constrained the coupling constant 
$g<3.2\times 10^{-8} \,\rm{GeV^{-1}}$ \cite{Ballou2015_OSQAR} for $m_\phi< 200\,\mu$eV. 
The sensitivity of the proposed experiment in comparison to that of
Ref.~\cite{Sikivie2014LCcircuit} is determined by the magnitude of the induced currents. Figure~\ref{fig:sensitivty} 
shows that there is approximately a 4 order of magnitude difference between the different proposals. We assume that $Q=10^4$ and $T=0.5\,\rm{mK}$ and choose $N_\ell=N_d=1$. Noting Ref.~\cite{Sikivie2014LCcircuit}
 we suppose the circuit would have a thermal noise of $\Delta\nu=4\,\rm{mHz}$, 
 $L_m=2.5\,\rm{\mu H}$, $L_c=0.5\,\rm{\mu H}$, $r_d=1\,\rm{cm}$, $a_m=7.4\times 10^{-4}\,\rm{m}$, 
 and $l=15\,\rm{cm}$. 
 The dimensions of the electric field are $r_e=30\,\rm{cm}$, $d=10\,\rm{cm}$, and $E_o=75 \,\rm{kV/cm}$, which are based on the dimensions from the 1/5-scale test studies in Ref.~\cite{osti_1261310} 
scaled-up to the dimensions 
of the neutron EDM experiment under development at Oak Ridge~\cite{nEDM:2019SNS}. 
To realize an electric field of that strength, the experiment would need to be 
performed in liquid $^4$He; the largest electric field established in vacuum is
$30\,\rm kV/cm$ \cite{Golub_30_Efield,osti_1261310}. 
Our estimated limits are shown in Fig.~\ref{fig:sensitivty} for 
the frequency window studied in Ref.~\cite{Sikivie2014LCcircuit}. 
Comparing the sensitivity of the limits proposed by Ref.~\cite{Sikivie2014LCcircuit} 
to those found by Ref.~\cite{ADMX-SLIC2020}, executed at $4.2\,{\rm K}$, and supposing a similar loss of sensitivity in the current case, we  see that values of the couplings smaller than 
the OSQAR limits can still be probed by this proposed set-up. 
We emphasize that the electric field studies for the Oak Ridge experiment are performed at a temperature of $300\, {\rm mK}$; at that temperature the proposed 
LC circuit experiment would lose about a factor of
20 in sensitivity to the scalar-photon coupling constant. 
We suppose further improvements
to our proposal 
could be made in differing 
ways, such as, e.g., 
by employing a slowly varying electric field to 
search for a beat frequency from the dark matter signal, or using more loops $N_\ell$.
We can also imagine using an array of SQUID or atomic magnetometers, as employed in neutron EDM
experiments, to make magnetic field measurements to set limits on ultralight scalar
dark matter directly, where we note that 
Oak Ridge experiment 
plans to use exceptionally large electric 
fields~\cite{nEDM:2019SNS}. 
For reference, we note an analogous study 
of the axion-gluon coupling from the PSI neutron EDM experiment~\cite{Abel2017axion_EDMPRX}. 
Finally, we note the development and demonstration 
of a quantum-enhanced sensor of
mechanical displacement and weak electric fields in a trapped
ion crystal within the $10 \,{\rm kHz}\, - \,10\,{\rm MHz}$ 
frequency range~\cite{Gilmore2021quantumenhanced};
such a scheme could potentially 
be sensitive to the scalar and axial vector dark
matter candidates we have noted here, 
as well as to the hidden photon and axion candidates 
they consider. 

{\it Summary.}
We have explored 
the modification of electrodynamics in the presence of ultralight dark matter 
of even intrinsic parity to show how, in the context of a resonant, superconducting LC circuit with
a large, static electric field, this can yield a new and sensitive direct probe of its existence
in the sub-$\mu$eV regime. In so doing 
we have exploited the large electric fields under development for neutron
EDM experiments, making dark matter probes of greatly 
enhanced sensitivity possible within the scope of 
existing technology. 

\begin{acknowledgments}
{\it Acknowledgements.}
We acknowledge 
partial support from the U.S. National Science Foundation under Award Number PHY-1950795 (UK REU) and 
the U.S. Department of Energy under contracts DE-FG02-96ER40989 and 
DE-SC001462. 
We thank P.~Sikivie for clarifying correspondence regarding Ref.~\cite{Sikivie2014LCcircuit}. 
\end{acknowledgments}

\bibliography{DM_LC_circuits.bib}% Produces the bibliography via BibTeX.

\end{document}